\documentclass[aps, prd, twocolumn, a4paper]{revtex4-1}
\usepackage{hyperref}
\usepackage{graphicx}
\usepackage{amsmath}
\usepackage{xcolor}
\usepackage{float}
\newcommand{\be}{\begin{equation}}
\newcommand{\ee}{\end{equation}}
\newcommand{\ba}{\begin{eqnarray}}
\newcommand{\ea}{\end{eqnarray}}
\newcommand{\nn}{\nonumber\\}

\begin{document}
	\title{Energy loss of heavy quarks in the isotropic collisional hot QCD medium}
	\author{Mohammad Yousuf Jamal $^{a,b}$}
	\email{mohammad.yousuf@niser.ac.in}
	\author{Vinod Chandra $^{a}$}
	\email{vchandra@iitgn.ac.in}
	\affiliation{$^a$Indian Institute of Technology Gandhinagar,  Gandhinagar-382355, Gujarat, India}
	\affiliation{$^b$School of Physical Sciences, National Institute of Science Education and Research, HBNI, Jatni-752050, India}

	\begin{abstract}

Collisional energy loss of heavy partons (charm and bottom quarks) has been determined within the framework of semi-classical transport theory implying Bhatnagar-Gross-Krook (BGK) collisional kernel. Hot QCD medium effects have been incorporated while employing a quasi-particle description of the medium in terms of effective gluons, quarks and antiquarks with respective temperature dependent effective fugacities. The momentum dependence of the energy loss for charm and bottom quark has been investigated. It is observed that with the increase in momentum of the heavy quarks, the loss increases sharply for the smaller values and reaches a saturation later. Further, as compared to the charm quark, bottom quark losses less energy at a particular momentum and collisional frequency. The energy loss is seen to increase with increasing collisional frequency. We also provide a comparative study of the results obtained using BGK-kernel than those using relaxation time approximation (RTA)-kernel and found them consistent with each other. The medium effects in all the situations are seen to play quite significant role.		
	\end{abstract}

\maketitle

	\section{Introduction}
The quark-gluon plasma (QGP) produced in the Relativistic Heavy-Ion Collider (RHIC) at Brookhaven National Laboratory and Large Hadron Collider (LHC) at CERN provides an opportunity to study the universe at the age of a few microseconds as well as the different phases of quantum chromodynamics (QCD). The QGP is seen to behave more like a near-perfect fluid (a tiny value of $\eta/s$)
~\cite{expt_rhic, expt_lhc, Heinz:2004qz}.
There have been several indirect probes proposed for the QGP in heavy-ion collision (HIC).
Among them, collective flow, jet quenching, quarkonia suppression and the suppression of high $p_T$ hadrons are the most reliable ones indicating the creation of the QGP.
The suppression of high $p_T$ hadrons is mainly caused by the energy loss of moving heavy partons in the QGP medium~\cite{Gyulassy:1999zd, Zakharov:2000iz, Djordjevic:2008iz, Baier:2001yt, Jeon:2003gi, Roy:2008tj}. This sets the motivation for the present work.

 Let us now discuss the current understanding of energy loss due to a fast charge particle in the interacting plasma medium.
In classical electrodynamics, the energy loss of a fast-moving charged particle, passing through the plasma is of particular importance in which one can relate the stopping power to the dielectric permittivity of the medium ~\cite{ichimaru1973basic}. In QCD, the analogue of this problem is the energy loss of high energy partons moving in the hot QCD medium. The high energy partons/heavy quarks that are created in the initial hard scatterings, in ultra-relativistic heavy-ion collisions, pass through the hot and dense matter produced after the collision and lose their energy through various processes ( collisions, radiation, {\it etc} ).
	  	  Some of the initial investigations primarily involve work by Bjorken who studied the collisional energy loss suffered by the high energy partons due to the elastic scatterings off thermal quarks and gluons in QCD plasma~\cite{bjorken1982energy}. Later on, Thoma and Gyulassy developed a formalism ~\cite{Thoma:1990fm} in which they obtained the collisional energy loss in terms of the longitudinal and transverse dielectric functions. In this approach, the infrared divergence is self-regulated due to the collective plasma effects. Within the finite temperature field theory approach, Braaten and Thoma ~\cite{Braaten:1991jj} had constructed a systematic framework of the energy loss for both soft and hard momentum transfers ~\cite{Mrowczynski:1991da, Thomas:1991ea, Koike:1992xs}. As the (momentum) anisotropy is present in all the stages of the system expansion, the authors in Refs.~\cite{Romatschke:2004au, Baier:2008js, Carrington:2015xca} have studied the anisotropic effects in the context of heavy quarks energy loss.
	      Apart from that, there are several excellent articles in which authors have discussed the energy loss of heavy quarks either through radiative or collisional means~\cite{Baier:2000mf, Jacobs:2004qv, Armesto:2011ht, Majumder:2010qh, Mustafa:1997pm, Dokshitzer:2001zm, Djordjevic:2003zk, Wicks:2007am, Abir:2011jb, Qin:2007rn, Cao:2013ita, Mustafa:2003vh, DuttMazumder:2004xk, Meistrenko:2012ju,
	  	Burke:2013yra, Peigne:2007sd, Neufeld:2014yaa, Chakraborty:2006db, Adil:2006ei, Peigne:2005rk}.
	  Recently, the polarization energy loss of heavy quarks while considering the hot viscous quark-gluon plasma has also been studied by Jiang et al.~\cite{Jiang:2014oxa, Jiang:2016duz}.
	  The heavy quarks collisional energy loss inside the quark-gluon plasma medium within the framework of transport approach, employing the finite RTA has been studied in Ref.~\cite{Elias:2014hua}. Same with the BGK collisional kernel has been investigated in Ref.~\cite{Han:2017nfz}.
			
	Here, we present the study of energy loss of bottom and charm quarks traversing through the isotropic collisional QGP within the effective transport theory approach. The collisions have been modelled using BGK collisional kernel in the Boltzmann Vlasov transport equation. Whenever a charged quark passes through the hot QCD medium, it induces the chromo-electric field that generates the Lorentz force which acts back on the moving quark. Hence, the incident quark loses its energy. The induced chromo-electric field is obtained in terms of longitudinal and transverse parts of medium dielectric permittivity which, in turn, expressed in terms of the gluon selfenergy. 
	    The gluon selfenergy has been obtained using the transport theory approach in the Abelian limit. This matches with the one-loop results from the HTL effective theory~\cite{Blaizot:1993be, Kelly:1994dh, Jamal:2017dqs}. In this limit, one needs to consider the high temperature where the perturbation theory is relevant. To incorporate the non-ideal hot QCD medium interaction effects in the analysis, the effective fugacity quasi-particle model (EQPM) ~\cite{chandra_quasi1, chandra_quasi2} has been employed which has been recently studied in Refs. \cite{Kurian:2019nna, Jamal:2018mog, Agotiya:2016bqr}. The results are then compared with those obtained while considering the ideal case (or leading order (LO)). A systematic comparison of the results on energy loss of heavy quarks employing BGK and RTA kernels (collisions) have been presented here.	
  
    The paper is organized as follows, in section~\ref{sec:EL}, we shall discuss the energy loss of heavy quarks moving in the hot QCD medium. Here, we shall obtain its expression in terms of dielectric functions using BGK-kernel within semi-classical transport theory approach. It is important to note that, in our previous work ~\cite{Kumar:2017bja}, we have already calculated the gluon selfenergy using BGK-kernel within the same method. We shall use a few of the results directly from there. In sections~\ref{el:RaD}, we shall discuss the various outcomes and provide a comparative analysis. Section~\ref{el:SaF}, is dedicated to the summary and future possibilities of the present work.
	
	\section{Energy loss of a moving heavy parton}
	\label{sec:EL}
	
	The motion of a classical color charged particle traversing through the chromodynamic field of plasma can be described by Wong's equations ~\cite{Wong:1970fu, Carrington:2015xca}. These equations are simply a set of classical equations of motion for a point-like particle interacting with a chromo-dynamical field, which is in the Lorentz covariant form given by,
	\ba
	\frac{dx^{\mu}(\tau)}{d\tau} &=& u^{\mu}(\tau),
	~~~~~~~\frac{dp^{\mu}(\tau)}{d\tau} =g q^{a}(\tau)F^{\mu\nu}_{a}(x(\tau))u_{\nu}(\tau),\nn
	\frac{dq^{a}(\tau)}{d\tau} &=& -gf^{abc}u_{\mu}(\tau)A^{\mu}_{b}(x(\tau))q_{c}(\tau),
	\label{eq:wong}
	\ea  
where $q^a(\tau)$ is the quark's color charge, $g$ is the coupling constant and  $F_{a}^{\mu\nu}$ is the chromodynamic field strength tensor. $\tau$,  $x^{\mu}$ and $p^{\mu}(\tau)$ are the proper time, trajectory and four momentum of the parton, respectively. The four velocity is  $u^{\mu}=\gamma(1,{\bf u})=\frac{p^{\mu}(\tau)}{m}$ ( where, $m$ being the mass of the particle). Here, we have $N_{c}^{2}-1$ chromo-electric/magnetic fields which belong to the $SU(N_c)$ gauge group, $A^{\mu}_{a}$ is the four potential.
	The expression of the energy loss can be obtained from the Wong's equations (given in Eq.~\ref{eq:wong}), following two assumptions. First, considering the gauge condition $u_{\mu}A^{\mu}_{a} = 0$ which says that $q^{a}$ is independent of $\tau$ and second, the quark's momentum and energy evolve in time without changing the magnitude of its velocity while interacting with the chromodynamic field. Now considering the zeroth component, $\mu =0$, in the second Wong's equation (Eq.~(\ref{eq:wong})) one can obtain the energy loss per unit length as,
	\ba
	-\frac{\text{dE}}{\text{d{\bf x}}}=g~q^a~\frac{{\bf u}}{|{\bf u}|}\cdot {\bf E}^{a}(X).
	\label{eq:el1}
	\ea
	The QGP is, in fact, a statistical system and hence, the polarization (${\bf E}^{a}_{ind}(X)$ ) and fluctuating (${\bf E}_{fl}^{a}$) chromo-electromagnetic fields produced at the same time when a heavy quark travels through the QGP.
	The ${\bf E}^{a}_{ind}(X)$  relates to the external current of the moving heavy
	quark whereas the randomly fluctuating $E_{fl}^{a}$, vanishes on the statistical averaging, {\it i.e.,} $\left < {\bf E}_{fl}^{a}\right > =0$. Therefore, the main contribution to the energy loss comes from the polarization of chromo-electric field that can be expressed as in ~\cite{Thoma:1990fm, Koike:1992xs, Elias:2014hua, Han:2017nfz},  
	\ba
	-\frac{\text{dE}}{\text{d{\bf x}}}=g~q^a~\frac{{\bf u}}{|{\bf u}|}\cdot {\bf E}_{ind}^{a}(X).
	\label{eq:el}
	\ea
	It is to be noted that the other components ($\mu =1,2,3$) are also important to consider while doing the full analysis on the energy loss ( such as energy loss due to fluctuations and correlations, {\it etc}). Here, we are mainly focusing on polarization energy loss. Therefore, it is sufficient to consider the zeroth component.
	
	To obtain the induced chromo-electric field, we start with the classical Yang-Mills equation in the Lorentz covariant form  given as,
	\ba
	\partial_{\mu}F_{a}^{\mu\nu}(X) =  J^{\nu}_{a, ind}(X) + J^{\nu}_{a, ext}(X).
	\ea
	Rewriting above equation in Fourier space, we obtain
	\ba
	-iK_{\mu} F'^{\mu\nu}_{a}(K) = J'^{\nu}_{a, ind}(K) + J'^{\nu}_{a, ext}(K),
	\label{eq:max}
	\ea
	where, $K\equiv K^{\mu} =(\omega,{\bf k })$. Now, the induced current, $J'^{\mu}_{a, ind}(K)$ in the Fourier space can be obtained as,
	\ba
	J'^{\mu}_{a, ind}(K) = \Pi^{\mu\nu}(K)A'_{\nu, a}(K). 
	\label{eq:linear induced current1}
	\ea
	Where, $\Pi^{\mu\nu}(K)$ is the gluon selfenergy (or the gluon polarization tensor).
	Using Eq.~(\ref{eq:max}) and Eq.~(\ref{eq:linear induced current1}) we get,
	\ba
	\bigg[K^{2} g^{\mu\nu} - K^{\mu} K^{\nu} + \Pi^{\mu\nu}(K)\bigg]A'_{\mu,a}(K) = - J'^{\nu}_{a, ext}(K).\nn 
	\label{eq:ext current1}
	\ea
	Considering the temporal axial gauge defined by $A_0=0$ ( with $ A^{j}_{a} = \frac{E^{j}_{a}}{i\omega}$), we can
	write Eq.~(\ref{eq:ext current1}) in terms of a chromo-electric field as, 
	\ba
	\bigg[(k^2-\omega^2)\delta^{ij} - k^{i} k^{j} + \Pi^{ij}(k)\bigg] E^{j}_{a}(K) = i \omega J'^{i}_{a, ext}(K).\nn
	\ea
	Rewriting the above equation,
	\ba
	[\Delta^{ij}(K)]^{-1} E^j_{a}(K) = i \omega J'^{i}_{a, ext}(K),
	\label{eq:propagator11}
	\ea
	or
	\ba
	E^{j}_{a}(K) = i \omega \, \Delta^{ij}(K) J'^{i}_{a,ext}(K),
	\label{eq:fi}
	\ea
	with 
	\ba
	[\Delta^{ij}(K)]^{-1} = (k^2-\omega^2)\delta^{ij} - k^{i} k^{j} + \Pi^{ij}(K),
	\label{eq:propagator1}
	\ea
	where $\Delta^{ij}(k)$ is the gluon propagator for the isotropic hot QCD medium. It is important to note that inclusion of collisions do not change the above expressions (only the form of  $\Pi^{ij}(K)$ modifies with the effects of collisions that we shall discuss in the next section). 
	The external current, ${\bf J}_{ext}^a(X)$ of a color point charge is given as,
	\ba
	{\bf J}_{ext}^a(X)= g~ q^a ~{\bf u}~\delta ^3 \left({\bf x}-{\bf u}t\right).
	\label{eq:cu}
	\ea
	In the  Fourier space ${\bf J_{ext}}^a(X)$ reads as,
	\ba
	{\bf J'}_{ext}^{a}\left(K\right)=\frac{i~ g q^a ~{\bf u}}{\omega -{\bf k}.{\bf u}+i 0^+}.
	\ea
	Here, we are considering a very near equilibrium situation. Therefore, all the collective modes are damped and the only stationary contribution	coming from the pole of ${\bf J'}_{ext}^{a}\left(K\right)$. Next, for the isotropic collisional case, the gluon selfenergy, $\Pi ^{\text{ij}}(K,\nu)$ relates with the dielectric permittivity, $\epsilon^{ij}(K,\nu)$ as,
	\ba
	\epsilon^{ij}(K,\nu)=\delta^{ij}-\frac{1}{\omega^2}\Pi ^{\text{ij}}(K,\nu),
	\label{eq:ep}
	\ea
	where $\nu$ is the collisional frequency. 
	
	The permittivity tensor can be expanded in terms of longitudinal and transverse components as,
	\ba
	\epsilon^{\text{ij}}(K,\nu)= A^{ij}~\epsilon_T(K,\nu)+B^{ij}~\epsilon_L(K,\nu),
	\label{eq:eplt1}
	\ea
	with
	\ba
	A^{\text{ij}}=\delta ^{\text{ij}}-\frac{k^i k^j}{k^2}, ~~~~~B^{\text{ij}}=\frac{k^i k^j}{k^2},
	\ea
	Using Eq.~(\ref{eq:propagator1}),~(\ref{eq:ep}) and ~(\ref{eq:eplt1}) in ~(\ref{eq:fi}), and taking the inverse Fourier transformation, the induced chromo-electric field in the coordinate space is obtained as,
	\ba
	{\bf E}^{a}_{ind}(X) &=&-i\frac{gq^{a}}{\pi}\int {d}\omega {d}^3{\bf k}\frac{1}{\omega~ k^2}\bigg[{\bf k}~({\bf k}\cdot{\bf u})\left(\epsilon^{-1}_L-1\right)\nn
	&+& \Big({k}^2 {\bf u}-{\bf k}~({\bf k}\cdot{\bf u})\Big) \bigg\{\left(\epsilon_T-\frac{k^2}{\omega^{2}}\right)^{-1}\nn
	&-&\left(1-\frac{k^2}{\omega^2}\right)^{-1}\bigg\}\bigg]\frac{e^{ i({\bf k}\cdot{\bf x}-\omega t)}}{\omega -{\bf k}\cdot{\bf u}+i 0^+}.
	\label{eq:inelc}
	\ea
	Integrating over, $\omega$ in Eqs.~(\ref{eq:inelc}) and substituting in Eq.~(\ref{eq:el}), we obtain the energy loss of a heavy parton moving in the hot QCD medium as,
	\ba
	-\frac{\text{dE}}{\text{d {\bf x}}}&=&\frac{C_F \alpha _s}{2 \pi ^2 |{\bf u}|}\int^{k_\infty}_{k_0} d^3{\bf k}\frac{\omega }{k^2}\bigg\{\left(k^2 |{\bf u}|^2-\omega ^2\right)\text{Im}(\omega ^2 \epsilon_T-k^2)^{-1}\nn
	&+&\text{Im}\epsilon^{-1}_{L}\bigg\}_{\omega ={\bf k \cdot u}},
	\label{eq:de}
	\ea
	where $\alpha_{s}(T)$ is the QCD running coupling constant at finite temperature~\cite{Laine:2005ai}
	and $C_F =4/3$ is the Casimir invariant in the fundamental representation of the $SU(3)$.
	
	Now to solve Eq.~(\ref{eq:de}), we need to know the form of the transverse and longitudinal components of the dielectric permittivity.
	We shall discuss this while considering the collisional effects using BGK-kernel in the next subsection.  
	
	\subsection{Dielectric permittivity in the presence of collisions}
	As mentioned earlier $\epsilon^{ij}(K,\nu)$ can be obtained within the semi-classical transport theory approach. To do so, one first needs to calculate the gluon selfenergy, $\Pi^{ij}(K,\nu)$. The detailed calculations of $\Pi^{ij}(K,\nu)$ considering BGK-collision kernel is provided in our previous work shown in Ref.~\cite{Kumar:2017bja}. There we have given a full calculation of gluon selfenergy for collisional anisotropic hot QCD medium that can be easily transferred to the isotropic collisional case by equating the anisotropic parameter, $\xi$ to zero. For the sake of completeness, we shall briefly provide its derivation.
	\begin{figure*}[ht]
		\centering
		\includegraphics[height=5cm,width=4.70cm]{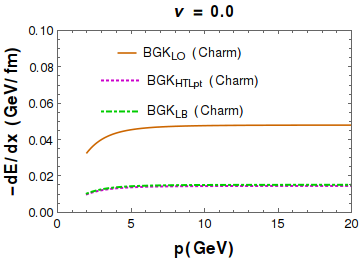}
		\includegraphics[height=5cm,width=4.70cm]{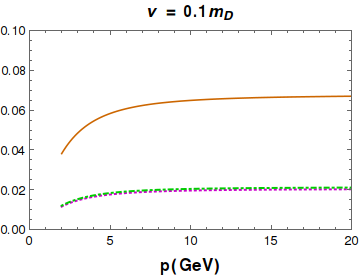}
		\includegraphics[height=5cm,width=4.70cm]{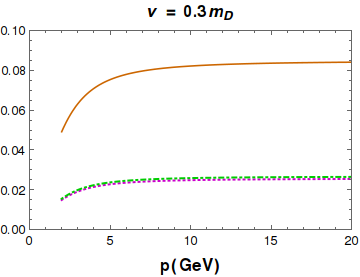}
		\caption{Energy loss of Charm ($m=1.8~GeV$) quark  for various values of $\nu$ at $T = 2 T_c$ and different EoSs. }
		\label{fig:charm}
	\end{figure*}
	\begin{figure*}[ht]
		\centering
		\includegraphics[height=5cm,width=4.70cm]{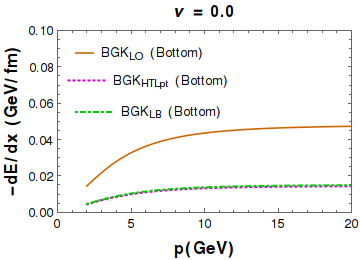}
		\includegraphics[height=5cm,width=4.70cm]{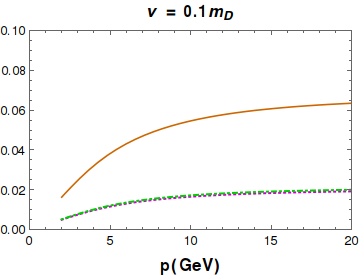}
		\includegraphics[height=5cm,width=4.70cm]{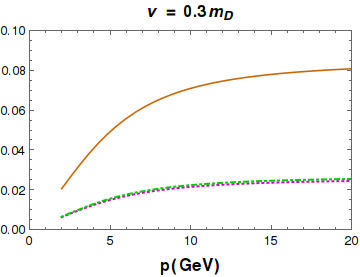}
		\caption{Energy loss of Bottom ($m=4.5~GeV$) quark  for various values of $\nu$ at $T = 2 T_c$ and different EoSs. }
		\label{fig:bottom}
	\end{figure*}
	We begin with the consideration that the current is induced in the plasma due to a slight deviation, $\delta f^a({\bf p}, X)$ in the particles distribution function from the equilibrium distribution function, $f^{0}({\bf p})$ such that,
	$	f^{0}({\bf p}) \gg \delta f^a({\bf p}, X)$.
	The induced current then could be obtained as,
	\ba
	J^{\mu,a}_{ ind}(X) = g \int \frac{d^{3}{\bf p}}{(2\pi)^{3}E}  p^{\mu}\delta f^{a}({\bf p},X),
	\label{eq:ind}
	\ea
	where
	\ba
	\delta f^{a}({\bf p},X) = 2 N_{c} \delta f_{g}^{a}({\bf p},X) + N_{f} (\delta f_{q}^{a}({\bf p},X) - \delta f_{\bar{q}}^{a}({\bf p},X)).\nn\ea
	Where $\delta f_{g}^{a}({\bf p},x)$, $\delta f_{q}^{a}({\bf p},x)$ and $\delta f_{\bar{q}}^{a}({\bf p},x)$ 
	are the fluctuating parts of the gluon, the quark and anti-quark densities, respectively. 
	In the Abelian limit, the fluctuation in the distribution function of each species in the medium  can be understood from 
	the Boltzmann-Vlasov ~\cite{Mrowczynski:1993qm, Romatschke:2003ms, Schenke:2006xu, Jiang:2016dkf} transport equation below,
	\ba
	u^{\mu }\cdot \partial^{X}_{\mu}\delta f_a^i(p,X)+g\theta _{i} u_{\mu }F^{\mu \nu }_a(X)\partial^p _{\nu }f^{i}(p)=C^i_a(p,X),\nn
	\label{eq:Vlasov}
	\ea
	where index -$i$ represents plasma species (quark, anti-quark and gluon).
	$\theta_{i}\in\{\theta_g,\theta_q,\theta_{\bar{q}}\}$ and have the values 
	$\theta_{g}=\theta_{q}=1$ and $\theta_{\bar{q}}=-1$. $C^i_a(p,X)$ is the collisional kernal which describes the effects of collisions between
	hard particles in the hot QCD medium. Here, we are initially focusing on $C^{i}_a(p,X)$ to be BGK-type~\cite{Bhatnagar:1954,Jiang:2016dkf,Schenke:2006xu}, which is defined as, 
		\ba
	C^{i}_a(p,X)=-\nu\left[f^{i}_a(p,X)-\frac{N^{i}_a(X)}{N^{i}_{\text{eq}}}f^{i}_{\text{eq}}(|\mathbf{p}|)\right]\,\text{,}\label{collision1}
	\ea
	where, $\nu$, is the collisions frequency. The BGK collision term \cite{Bhatnagar:1954} describes the equilibration of the system, due to the collisions, in a time proportional to $\nu^{-1}$. Here, we are assuming $\nu$ to be independent of momentum and particle species. The collision term, $C^{i}_a(p,X)$ in the special case, $\frac{N^{i}_a(X)}{N^{i}_{\text{eq}}}\rightarrow 1$ with $\nu=\frac{1}{\tau}$, $\tau$  being relaxation time, gives the form of the RTA kernel. BGK modelling is comparatively more reliable in the sense that it conserves the particle number instantaneously which is absent in RTA approach, {\it i.e.,} while using BGK kernel we have,
	\ba
	\int \frac{d^{3}p}{(2\pi)^3}C^{i}_a(p,X)=0.
	\ea 
	Now, the particle number, $N^{i}_a(X)$ and its equilibrium value, $N^{i}_{\text{eq}}$ are defined as follows,
	\ba
	N^{i}_a(X)=\int \frac{d^{3}p}{(2\pi)^3} f^{i}_a(p,X)\text{ , ~} \\ N^{i}_{\text{eq}} = \int \frac{d^{3}p}{(2\pi)^3} f^{i}_{\text{eq}}(|\mathbf{p}|) = \int \frac{d^{3}p}{(2\pi)^3} f^{i}(\mathbf{p})\text{.}
	\ea
	\begin{figure}[ht]
		\centering
		\includegraphics[height=5cm,width=7.0cm]{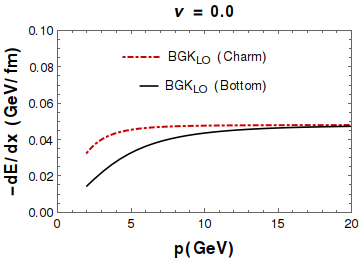}
		\includegraphics[height=5cm,width=7.0cm]{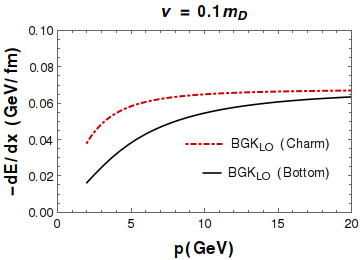}
		\caption{Energy loss of charm and bottom quark at $T =2 T_c$ and $\nu =0.0~ \text{and}~ 0.1 m_{D}$ for leading order case.}
		\label{fig:CB}
	\end{figure}
	Next, solving Eq.~(\ref{eq:Vlasov}) for $\delta f^{a}_{q, {\bar q}, g}$ in the Fourier space and using Eq.~(\ref{eq:ind}) and Eq.~(\ref{eq:linear induced current1}), we obtained the spatial components of gluon selfenergy, $\Pi^{ij}(K)$ as,
	\ba
	\Pi ^{ij}(K,\nu)&=& m_D^2(T)\int \frac{d\Omega }{4 \pi }u^{i} u^{l}\left\{u^{j} k^{l}+\left(\omega -{\bf k}\cdot {\bf u}\right)\delta ^{lj}\right\}\nn
	&\times& D^{-1}\left(K,\nu \right),
	\label{eq:pi}
	\ea
	where
	\ba
	D\left(K,\nu \right)&=&\omega +i \nu-{\bf k}\cdot {\bf u}.
	\ea
	The Debye mass, $m_D$ is given as,
	\ba
	\label{dm}
	m_D^2&=& -4 \pi \alpha_{s}(T) \bigg(2 N_c \int \frac{d^3 p}{(2 \pi)^3} \partial_p f_g ({\bf p})\nn
	&+& 2 N_f  \int \frac{d^3 p}{(2 \pi)^3} \partial_p f_q ({\bf p})\bigg).
	\ea
	Here, we are adopting quasi-particle distribution functions considering the EQPM,  $ f_{eq}\equiv \lbrace f_{g}, f_{q} \rbrace$ 
	that describes the strong interaction effects in terms of effective fugacities, $z_{g,q}$~\cite{chandra_quasi1, chandra_quasi2},
	\ba
	\label{eq1}
	f_{g/q}= \frac{z_{g/q}\exp[-\beta E_p]}{\bigg(1\mp z_{g/q}\exp[-\beta E_p]\bigg)},
	\ea
	where, $E_{p}=|{\bf p}|$ for the gluons and, $\sqrt{|{\bf p}|^2+m_q^2}$ for the quark degrees of freedom ($m_q$, denotes the mass of the quarks).
	The fugacity parameter, $z_{g/q}\rightarrow 0$ as temperature $T\rightarrow \infty$.
	Since the model is valid only in the deconfined phase of QCD (beyond $T_c$), therefore, the mass of the light quarks can be neglected as compared to the temperature.
	Next, the $\Pi^{ij}(K,\nu)$ can be further decomposed (in the isotropic collisional case) in terms of its longitudinal and transverse parts as,
	\ba
	\Pi^{\text{ij}}(K,\nu)= A^{ij}~P_T(K,\nu)+ B^{ij}P_L(K,\nu),
	\label{eq:eplt}
	\ea
	where the structure constants, $P_T(K,\nu)$ and $P_L(K,\nu)$ for the isotropic collisional case can be obtained as, 
	\ba
	P_T(K,\nu) &=&\frac{m_D^2~\omega}{4 k^3} \bigg[2 k (\omega +i \nu ) + \Big(k^2+(\nu -i \omega )^2\Big)\nn
	&\times& \log \Big(\frac{\omega+i \nu +k }{\omega+i \nu -k }\Big)\bigg],
	\ea
	and
	\ba
	P_L(K,\nu) =-\frac{\omega ^2 m_D^2 \left(1-\frac{\omega +i \nu}{2 k}\log \left(\frac{\omega+i \nu +k }{\omega+i \nu -k }\right)\right)}{k^2 \left(1-\frac{i \nu}{2 k}  \log \left(\frac{\omega+i \nu +k }{\omega+i \nu -k }\right)\right)}.
	\ea
	From Eqs.~(\ref{eq:ep}),~(\ref{eq:eplt1}) and ~(\ref{eq:eplt}), one can obtain the longitudinal and transverse part of the dielectric permittivity, respectively as,
	\ba
	\epsilon_L(K,\nu)&=&1- \frac{P_L(K,\nu)}{\omega^2},
	\ea\ba
	\epsilon_T(K)=1- \frac{P_T(K,\nu)}{\omega^2},
	\ea
	that could be further written as,
	\ba
	\epsilon _L(K,\nu)=1+\frac{m_D^2 \left(2 k-(\omega +i \nu ) \log \left(-\frac{k+i \nu +\omega }{k-i \nu -\omega }\right)\right)}{k^2 \left(2 k-i \nu  \log \left(-\frac{k+i \nu +\omega }{k-i \nu -\omega }\right)\right)},
	\ea
	and
	\ba
	\epsilon _T(K,\nu)&=&1-\frac{m_D^2}{2~ \omega~k} \bigg[\frac{(\omega +i \nu )}{k}+\frac{1}{2}\left(1-\frac{(\omega +i \nu )^2}{k^2}\right)\nn
	&\times&\log \left(-\frac{k+i \nu +\omega }{k-i \nu -\omega }\right)\bigg].
	\ea
	As mentioned earlier, the RTA kernel can be obtained from the BGK one (given in Eq.~(\ref{collision1})) by considering $\frac{N^{i}_a(X)}{N^{i}_{\text{eq}}}\rightarrow 1$ and $\nu=\frac{1}{\tau}$, $\tau$  being relaxation time. We repeated the whole analysis using RTA term in Eq.~(~\ref{collision1}) and obtained the longitudinal and transverse part of the dielectric permittivity considering the hot QCD medium as,
	\ba
	\epsilon _L(K,\nu)=1+\frac{m_D^2~ \omega'}{k^2 \omega } \left(1-\frac{\omega'}{2 k} \log \left[\frac{\omega'+k}{\omega'-k}\right]\right),
	\ea
	and
	\ba
	\epsilon _T(K,\nu)=1-\frac{m_D^2}{\omega ^2} \left(\frac{\omega~  \omega'}{k^2}+\frac{\omega}{2 k}  \left(1-\frac{\omega'^2}{k^2}\right) \log \left[\frac{\omega'+k}{\omega'-k}\right]\right),\nn
	\ea 
	where $\omega'=\omega+i\nu$.
	Now, in both the cases, using $\epsilon _L(K,\nu)$ and $\epsilon _T(K,\nu)$ in Eq.~(\ref{eq:de}), one can obtain the energy loss for the heavy quarks (charm and bottom) moving in the isotropic collisional hot QCD medium. In the next section, we shall show and discuss the various plots regarding energy loss of charm and bottom against their momenta at different collisional frequencies.
	
	\section{Results and discussion}
	\label{el:RaD}
	\begin{figure}[ht]
		\centering
		\includegraphics[height=5cm,width=8.0cm]{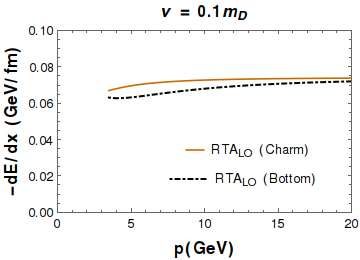}
		\caption{Energy loss of charm and bottom quark using RTA-kernel at $\nu = 0.1 m_{D}$ and  $T = 2 T_c$.}
		\label{fig:RCB}
	\end{figure}

	\begin{figure*}[ht] 
	\centering
	\includegraphics[height=6cm,width=8.60cm]{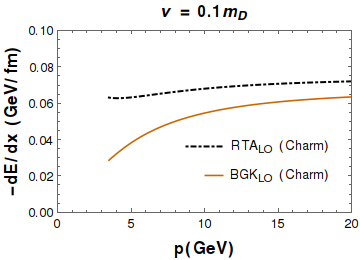}
	\includegraphics[height=6cm,width=8.60cm]{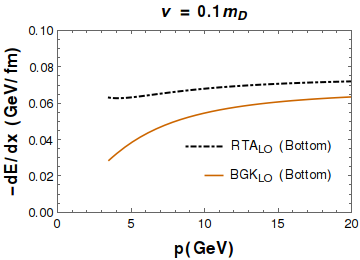}
	\caption{Comparison of energy loss of charm (left) and bottom (right) quark using BGK and RTA- kernels at $\nu=0.1 m_{D}$ and $T = 2 T_c$. }
	\label{fig:RBbottom}
\end{figure*}

The energy loss of heavy quarks (charm and bottom) moving in the isotropic collisional hot QCD/QGP medium has been studied. In this context, Eq.(\ref{eq:de}) have been solved numerically. To perform the numerical integration, the lower and upper limits have been, respectively taken as, $k^{EoSs}_{0}=0$ and $k^{EoSs}_{\infty}\sim m^{EoSs}_{D}(T)$ for each Equation of State (EoS). The results obtained for the ideal case (or the leading order (LO)) are compared with the non-ideal cases ( $(2+1)$ lattice EoS and 3-loop HTL EoS denoted as $LB$ and $HTL_{pt}$, respectively in the plots). The different collisional frequencies, $\nu$, have been chosen to investigate its impact on energy loss and also compared with collision-less case $\nu=0.0$. Few works while considering the LO case using BGK-kernel are already available in the literature ~\cite{Han:2017nfz}. Our numbers for the LO are slightly different. The reason is that in the present case, the coupling constant and the Debye mass are not fixed. Instead, they are temperature dependent. Here, we are working at temperature, $T=2~T_c$ where, $T_c=0.17 ~GeV$.

    In Fig.\ref{fig:charm} and Fig.\ref{fig:bottom}, energy loss of charm and bottom quarks have been plotted for collision-less case $\nu=0$ (left), with collision frequency, $\nu=0.1~m_{D}$ (center) and $\nu=0.3~m_{D}$ (right), respectively. We have noticed that the energy loss initially increases and then saturates with the increasing particle's momentum, which match with the results that are already present in the literature~\cite{Han:2017nfz}. While considering the non-ideal EoSs ($LB$ and $HTL_{pt}$) the energy loss (for both the heavy quarks) have been found suppressed as compared to the ideal one ($LO$) at fixed collisional frequency. Whereas, an increase in the collision frequency causes more energy loss. In Fig.\ref{fig:CB}, we compared the energy loss of charm and bottom quarks at $\nu=0.0 ~\text{and}~0.1~m_{D}$ for the leading order case and observed that as compared to bottom quark, charm quark loses more energy at fixed momentum. This supports the fact that heavy particle loses less energy while moving in a medium than the lighter one, given the same conditions.
	
	As mentioned earlier, we also obtained the results using the RTA-collisional kernel to have a comparative study with the BGK case. Mathematically, the difference occurs only in the expressions of the longitudinal and transverse part of the dielectric tensor. The energy loss of charm and bottom quarks using RTA kernel has been plotted in Fig.~\ref{fig:RCB} and observed the same patterns as the BGK one at $\nu = 0.1m_{D}$. In Fig.~\ref{fig:RBbottom}, a comparison between RTA and BGK results have been shown for charm quark (in the left panel) and a bottom quark (in the right panel) at $\nu = 0.1m_{D}$. It has been observed that, given the same momentum and the collisional frequency, the energy loss is seen more in RTA case as compared to the BGK one.

	\section{Summary and Future aspects}
	\label{el:SaF}
The energy loss of the heavy quarks moving through the isotropic collisional hot QCD medium produced in the relativistic heavy-ion collision experiments has been investigated. The expression of energy loss is obtained in terms of the longitudinal and transverse part of the dielectric permittivity. Employing the effective kinetic theory in the high-temperature limit (considering the Abelian part) using the BGK-kernel, gluon selfenergy and dielectric permittivity tensor have been obtained. We found that the energy loss increases initially with the momentum and then saturates. The energy loss also found to be greater for higher collisional frequency.
Moreover, the bottom quark (more massive quark) is found to lose less energy than the charm quark (lighter quark) for the same collisional frequency and momentum. We also performed the same analysis considering the RTA-kernel and provided a comparative study. It has been observed that the expression of dielectric permittivity modified and a slight deviation has been found in the results. Considering the same values of momentum and collisional frequency, more energy loss is observed in the RTA case than the BGK one. 

We intend to incorporate the momentum anisotropy in the formalism soon in the near future. The inclusion of viscous effects by employing the effective quasi-particle picture will also be an immediate extension to the present work. In addition, $R_{AA}$ would be another important quantity to investigate as it is essential to relate the theoretical estimations with the experimental observations.
    
	\section{Acknowledgements}
 M. Y. Jamal would like to thank Prof. Jitesh R. Bhatt and Dr. Avdhesh Kumar for fruitful discussions and valuable inputs that helped in making the present manuscript better. M. Y. Jamal further acknowledges NISER Bhubaneswar for providing postdoctoral position. V. Chandra would like to sincerely acknowledge DST, Govt. of India for Inspire Faculty Award -IFA13/PH-15 and Early Career Research Award(ECRA) Grant 2016. We would also like to acknowledge people of INDIA for their generous support for the research in fundamental sciences in the country.

\end{document}